\PassOptionsToPackage{final}{graphicx}
\documentclass[conference]{IEEEtran}

\newcommand{\bm}{\mathbf}
\newcommand{\be}{\begin{equation}}
\newcommand{\ee}{\end{equation}}
\newcommand{\bea}{\begin{eqnarray}}
\newcommand{\eea}{\end{eqnarray}}
\newcommand{\Lamb}{{\boldsymbol\Lambda}}
\newcommand{\x}{{\bm x}}

\newcommand{\br}{{\bm r}}

\newcommand{\bA}{{\bm A}}
\newcommand{\bG}{{\bm G}}
\newcommand{\bL}{{\bf L}}
\newcommand{\bR}{{\bm R}}

\newcommand{\bT}{{\bf T}}
\newcommand{\bU}{{\bf U}}

\newcommand{\bW}{{\bf W}}

\newcommand{\bd}{{\bf d}}

\newcommand{\I}{{\bm I }}

\newcommand{\bPhi}{\mbox{\boldmath{$\Phi$}}}

\usepackage{algorithm}
\usepackage{algpseudocode}

\usepackage{epsfig}
\usepackage{amsmath}
\usepackage{graphicx}
\usepackage{epsfig}
\usepackage{amsmath}
\usepackage{subfigure}
\usepackage{algorithm}
\usepackage{amssymb}
\usepackage{algpseudocode}
\algrenewcommand\algorithmicrequire{\textbf{Input:}}
\algrenewcommand\algorithmicensure{\textbf{Output:}}
\usepackage{multicol}
\usepackage{color}
\usepackage{cite}

\begin{document}

\title{Interference Localization for Uplink OFDMA Systems in Presence of CFOs}
\author{\normalsize Arman Farhang$^*$, Arslan Javaid Majid$^\dagger$, Nicola Marchetti$^*$, Linda E. Doyle$^*$ and Behrouz Farhang-Boroujeny$^\dagger$  
\\$^*$CTVR / The Telecommunications Research Centre, Trinity College Dublin, Ireland, \\
$^\dagger$ECE Department, University of Utah, USA. \\
Email: \{farhanga, marchetn, ledoyle\}@tcd.ie, \{majid, farhang\}@ece.utah.edu}
\maketitle

\begin{abstract}
Multiple carrier frequency offsets (CFOs) present in the uplink of orthogonal frequency division multiple access (OFDMA) systems adversely affect subcarrier orthogonality and impose a serious performance loss. In this paper, we propose the application of time domain receiver windowing to concentrate the leakage caused by CFOs to a few adjacent subcarriers with almost no additional computational complexity. This allows us to approximate the interference matrix with a quasi-banded matrix by neglecting small elements outside a certain band which enables robust and computationally efficient signal detection. The proposed CFO compensation technique is applicable to all types of subcarrier assignment techniques. Simulation results show that the quasi-banded approximation of the interference matrix is accurate enough to provide almost the same bit error rate performance as that of the optimal solution. The excellent performance of our proposed method is also proven through running an experiment using our FPGA-based system setup.

\end{abstract}

\section{Introduction}\label{sec:intro}
Because of its high spectral efficiency and robustness against multipath fading, orthogonal frequency division multiple access (OFDMA) technology has been adopted for the physical layer of several wireless standards, e.g., WiMAX and 3GPP Long Term Evolution (LTE). OFDMA systems have high sensitivity to synchronization errors especially in the uplink. The inevitable imperfect synchronization breaks the orthogonality between the subcarriers and causes multiple access interference (MAI) as well as self-user interference \cite{Chockalingam2009}. Synchronization errors due to the timing misalignment of the users' signals can be eliminated by a choice of long enough cyclic prefix (CP), to obtain a set of quasi-synchronous OFDM signals. The residual timing mismatch between different users will be reflected as part of their respective channel impulse responses and hence, can be compensated for at the equalization stage \cite{Moreli2007}. 

Because of the fact that the received signal at the base station (BS) is a superposition of all the users' signals, multiple CFOs appear in the received signal. This results in inter-carrier interference (ICI) and hence MAI. To remove MAI a two-step process is applied. First, an appropriate signal processing method is used to estimate the different users' CFOs. A CFO compensation method is then employed to remove MAI effects. Several algorithms have been proposed in the literature for the former \cite{Linesearch2007, Ghrayeb2009}, \cite{Moreli2007}. However, the focus of this paper is on the latter, i.e., the MAI removal methods.

To tackle the ICI problem caused by CFOs, in the uplink of OFDMA based systems, a number of solutions have been proposed in the literature \cite{Arsalan2007,Lataief2005,Cao2007,Hsu2008,Lee2011,Lee2012,S.Ahmed2009,AF2013,Chen2010}.
The authors in \cite{Arsalan2007} and \cite{Lataief2005} propose interference cancellation techniques to eliminate the effect of ICI and MAI using tentatively detected data symbols of different users in each iteration. Chen et al, \cite{Chen2010}, extend this study by suggesting a joint minimum mean square error frequency domain equalization (MMSE-FDE) and CFO compensation technique with interference cancellation. Ahmed and Zhang \cite{S.Ahmed2009} suggest a method of preconditioning the received signal vector, before applying the DFT, in order to limit the interfering subcarriers to a few adjacent ones only. This reduces the complexity of successive interference cancellation significantly. This method was originally proposed by Schniter, \cite{Schniter2004}, in the context of single user OFDM systems in time varying channels. The drawback of the interference cancellation solutions is that their performance degrades as the CFO increases \cite{Arsalan2007}, \cite{Lataief2005}, \cite{S.Ahmed2009}. Besides, such methods may suffer from the error propagation problem, since wrong decisions may be fed back for cancellation.

In another class of CFO compensation techniques, which is of more interest in this paper, the CFO effects are modelled as multiplication of an interference matrix to a composite data vector of all the users which forms a system of equations. Solving this system of equations gives all the users' data symbols, free of MAI and self-interference. However, the solution needs an inversion of the interference matrix whose size equals the total number of subcarriers, which, in practice, can be as large as a few thousands. This clearly makes the solution highly complex. Hence, solutions with reduced complexity have to be sought. Such a solution for the case where interleaved subcarrier allocation is used is proposed by Hsu and Wu \cite{Hsu2008}. In a more recent work, the authors take advantage of interference matrix structure in interleaved and block-interleaved subcarrier assignments to further reduce the computational cost \cite{AF2013}. Lee et al, \cite{Lee2012}, consider the generalized carrier allocation scheme (G-CAS) and proposed an MMSE compensation technique using conjugate gradient (CG) algorithm. This method reduces the computational complexity to a great extent. Another relevant work is that of Cao et al, \cite{Cao2007}, where the authors approximated the interference matrix by a banded one, simply, by assuming the elements that are off from the diagonal by a spacing of greater than $D$ (a design parameter) are equal to zero. It is then noted that this banded property of the (approximated) interference matrix can be used to find the desired solution with a low computational complexity that is of the order of $ND^2$.

In this paper, we propose a new approach for limiting the interference from each subcarrier to a few of its adjacent ones using a receiver windowing technique. This enables us to approximate the interference matrix to a quasi-banded one with a negligible loss in performance. This is in contrast to the solution in \cite{Cao2007} where the approximation of the original interference matrix to a banded one incurs a significant loss in performance. Our solution will result in a similar performance to that of \cite{Lee2012}, albeit at the cost of some loss (less than 10\%) in bandwidth efficiency; see the details in Section~\ref{sec:Rx_filtering}. Moreover, similar to \cite{Lee2012}, it is applicable to G-CAS. In addition, our novel solution, for typical cases of interest, reduces the computational complexity by over an order of magnitude when compared to the CG method; the most efficient algorithm in the literature.

The rest of the paper is organized as follows. In Section~\ref{sec:system_model} we formulate the CFO problem in the uplink of OFDMA based systems. Section~\ref{sec:Rx_filtering}, discusses our proposed MAI reduction and CFO compensation techniques in detail. The BER performance of the proposed technique and comparisons with existing methods are presented in Section~\ref{sec:Sim_results}. Results of an experimental study are presented in Section~\ref{sec:Experimental_Study}. A system complexity analysis that compares the proposed method with its counterparts from literature is presented in Section~\ref{sec:sys_comp} and finally, the conclusions of the paper are drawn in Section~\ref{sec:Conclusion}. 

Throughout the paper, matrices, vectors and scalar quantities are denoted by boldface uppercase, boldface lowercase and normal letters, respectively. $[\bA]_{m,n}$ represents the element in the $m^{\rm{th}}$ row and $n^{\rm{th}}$ column of matrix $\bA$, $\bA^{-1}$ signifies the inverse of $\bA$, $\I_M$ is the identity matrix of size $M$ and ${\bf 0}_{m\times n}$ is the zero matrix of size $m$ by $n$. The superscripts $(\cdot)^T$ and $(\cdot)^H$ indicate transpose and conjugate transpose of a matrix, respectively. Finally, ${\rm diag}(\x)$, $\ast$ and $|\cdot|$ represent a diagonal matrix with diagonal elements belonging to the vector $\x$, linear convolution and absolute value, respectively.

\section{System Model}\label{sec:system_model}
The uplink of an OFDMA system with $K$ users communicating with the base station through independent multipath wireless channels is considered. We assume to have a total number of $N$ subcarriers in each OFDM symbol. Therefore, each user can have $L=N/K$ subcarriers. In this paper, we consider the generalized subcarrier assignment scheme where randomly chosen subcarriers are assigned to different users. The $L\times1$ vector ${\bf{d}}^{(i)}$ contains data symbols of the $i^{\rm th}$ user. It is worth mentioning that the subcarriers of distinct users are mapped on mutually disjoint subsets of the available subcarriers. Hence, if the subcarriers allocated to the $i^{\rm th}$ and $j^{\rm th}$ users belong to the sets $\Psi_i$ and $\Psi_j$, respectively, then $\Psi_i\cap\Psi_j=\phi, i\neq j$ and $\bigcup^{K}_{m=1}\Psi_m=\{1,\ldots,N\}$. In the OFDMA transmitter, the first step is carrier allocation. Thus, the $i^{\rm th}$ user signal vector after subcarrier mapping is

\be\label{eqn:Si}
{\bf{s}}^{(i)}_{\rm f}={\bf{\Gamma}}^{(i)} \bd^{(i)},
\ee  
where the subscript $\rm f$ in ${\bf{s}}^{(i)}_{\rm f}$ stresses that the data is in frequency domain and ${\bf{\Gamma}}^{(i)}$ is $N{\times}L$ subcarrier allocation matrix of user $i$. Columns of ${\bf{\Gamma}}^{(i)}$ are equal to the columns of an $N \times N$ identity matrix that belong to $\Psi_i$.
The output of the IDFT unit for user $i$ is given by
\be\label{eqn:si}
{\bf{s}}^{(i)}_{\rm t}={\bf F}^{H}_{N}{\bf{s}}^{(i)}_{\rm f},
\ee
where ${\bf{F}}_N$ is an $N$-point DFT matrix with elements $[{\bf{F}}_N]_{n,k}={\frac{1}{\sqrt{N}}}{e^{\frac{-j2{\pi}nk}{N}}}$. The subscript $\rm t$ in ${\bf{s}}^{(i)}_{\rm t}$ emphasises that the vector is in time domain. 
Finally, the cyclic prefix (CP) and cyclic suffix (CS) whose lengths are $N_{\rm CP}$ and $N_{\rm CS}$ will be appended to the first and last part of the signal. In order to avoid self and multi-user interference due to the timing offsets of the users, a CP longer than both the maximum channel delay spread and the two way propagation delay is required and the residual timing errors will be incorporated in the channel impulse responses of the users; thereby, inter-symbol interference (ISI) between different users will be avoided \cite{Moreli2007}. Therefore, the cyclically extended signal can be shown as

\be\label{eqn:Sicp}
{\tilde{\bf{s}}}^{(i)}_{\rm t}={\bf{T}}{\bf{s}}^{(i)}_{\rm t},
\ee
where ${\bf{T}}=[\bG_{\rm CP}^T,\I_N^T,\bG_{\rm CS}^T]^T$ is the corresponding cyclic extension matrix and the rows of $\bG_{\rm CP}$ and $\bG_{\rm CS}$ matrices include the last $N_{\rm CP}$ and the first $N_{\rm CS}$ rows of the identity matrix $\I_N$, respectively. The wireless channels for different users are assumed to be statistically independent with respect to each other and time invariant during one OFDMA symbol. If the channel impulse response for user $i$ has the length equal to $N_{\rm ch}$ samples, it can be denoted by the vector ${\bf h}^{(i)}=[{h}^{(i)}_{0},\ldots,{h}^{(i)}_{N_{\rm ch}-1}]^T$ whose elements are assumed to be statistically independent complex Gaussian random variables. Considering the impact of CFOs from different users, the received signal at the receiver can be depicted by
\be\label{eqn:rcp}
{\tilde{\br}}=\sum^{K}_{i=1}{{\boldsymbol\Phi}(\epsilon_i)({\bf{h}}^{(i)}\ast{\tilde{\bf{s}}}^{(i)}_{\rm t})}+{\boldsymbol{\nu}},
\ee
where ${\boldsymbol\Phi}(\epsilon_i)$ is the $N_T{\times}N_T$ diagonal CFO matrix whose diagonal elements are $\{1,{e^{{\frac{j2{\pi}{\epsilon_i}}{N}}}},\ldots,{e^{{\frac{j2{\pi}{\epsilon_i}(N_T-1)}{N}}}}\}$, $\epsilon_i$ is the $i^{\rm{th}}$ user's CFO normalized by subcarrier spacing and $N_T=N+N_{\rm CP}+N_{\rm CS}$. Finally, $\boldsymbol\nu$ is the complex additive white Gaussian noise (AWGN) vector, i.e, $\boldsymbol\nu~{\sim}~CN(0,{{\sigma_\nu}^{2}}{\I_{N_T}})$ and ${\sigma_\nu}^{2}$ is the noise variance. 

\section{Filtering at the Receiver for MAI Reduction}\label{sec:Rx_filtering}
In conventional OFDMA systems, after CP removal, the received signal vector is passed through a DFT block. This is equivalent to analyzing each received OFDM symbol through a bank of filters that are characterized by a rectangular prototype filter. Such a filter bank system suffers from large side-lobes and consequently a significant level of MAI when different subcarriers are not synchronized in frequency with respect to one another. 

\begin{figure}[t]
		\centering
    \includegraphics[scale=0.52]{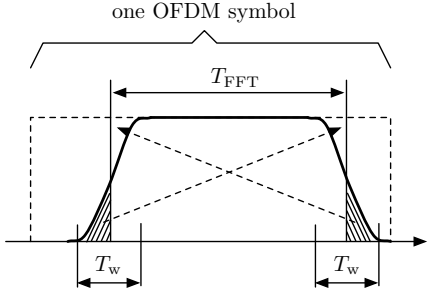}
    \caption{A raised-cosine window and the process of aliasing in time domain. The excess samples beyond the length $T_{\rm FFT}$ at each side are added to the attenuated samples at the other side.}

\label{fig:Rx_window}
\vspace{-2 mm}
\end{figure}

In order to reduce the ICI to a limited number of adjacent subcarriers and hence, limit the interference matrix to a quasi-banded one, we borrow the following idea from the discrete multi-tone (DMT) literature\footnote{DMT is the equivalent name for OFDM in the digital subscriber lines (DSL) literature.}. In \cite{Zipper}, to mitigate near-end cross-talk and radio frequency interference in very high bit-rate digital subscriber lines (VDSL), it has been proposed to replace the rectangular prototype filter/window in an OFDM receiver by a window with smooth roll-offs at the sides, as shown in Fig.~\ref{fig:Rx_window}. A raised-cosine window is proposed for this application in \cite{Zipper}. Given that the number of samples in the time domain are $N+N_{\rm w}$ and we need to analyze the signal samples in the frequency domain at $N$ equally spaced samples, one may conveniently alias the time domain signal, as shown in Fig.~\ref{fig:Rx_window}, and then apply an $N$-point DFT to the result \cite{Zipper}.

An analysis of the raised-cosine window that provides insight to its impact on side-lobe suppression is recently reported in  \cite{FarhangOFDMvsFBMC}. It is noted that if $T_{\rm FFT}$ denotes the length of DFT and $T_{\rm w}$ the duration of the roll-off at each side, the raised-cosine window can be mathematically expressed as 
\begin{equation}
g(t)={\rm rect}\left(\frac{t-T_{\rm FFT}/2}{T_{\rm FFT}}\right)*c(t),
\end{equation}
where 
\begin{equation}
c(t)=\frac{\pi}{2T_{\rm w}}\sin\left(\frac{\pi t}{T_{\rm w}}\right){\rm rect}\left(\frac{t-T_{\rm w}/2}{T_{\rm w}}\right),
\end{equation}
and ${\rm{rect}(\cdot)}$ is the rectangular function. Accordingly, in the frequency domain,
\begin{equation}
|G(f)|=T_{\rm FFT}|{\rm sinc}(fT_{\rm FFT})|\times |C(f)|,
\end{equation}
and
\begin{equation}
|C(f)|=\left|\frac{\cos(\pi fT_{\rm w})}{1-4f^2T_{\rm w}^2}\right|.
\end{equation}
A point to note here is that $|C(f)|$ has a `$\rm sinc$' shape with the main lobe of $3/T_{\rm w}$ wide. It also drops to below $-12$~dB beyond the frequency range $(-1.1/T_{\rm w},1.1/T_{\rm w})$. If this attenuation is taken as sufficient to suppress the subcarrier side-lobes (numerical results presented later shows this is a good compromise choice), one will find that the interference matrix becomes (with a good approximation) quasi-banded with a bandwidth of $2\lfloor 1.1T/T_{\rm w}\rfloor+1$, where $\lfloor\cdot\rfloor$ rounds down the number inside.

Since, the first $N_{\rm GI}$ samples of ${\tilde{\br}}$ in (\ref{eqn:rcp}) are affected by the channels of the users, we will discard them using guard interval removal matrix $\bR_{\rm GI}=[{\bf 0}_{(N+N_{\rm w})\times N_{\rm GI}},\I_{(N+N_{\rm w})}]$ and we have
\be\label{eqn:r}
\br=\sum^{K}_{i=1}{{e^{{\frac{j2{\pi}{\epsilon_i}N_{\rm GI}}{N}}}}{{\widetilde{\bPhi}}(\epsilon_i)\bar{\bT}{\bf{H}}^{(i)}_{\rm t}{\bf{s}}^{(i)}_{\rm t}}}+\bR_{\rm GI}{\boldsymbol{\nu}},
\ee
where ${\widetilde{\bPhi}}(\epsilon_i)={\rm diag}(1,{e^{{\frac{j2{\pi}{\epsilon_i}}{N}}}},\ldots,{e^{{\frac{j2{\pi}{\epsilon_i}(N+N_{\rm w}-1)}{N}}}})$, $\bar{\bT}=[{\bG}_{\rm W}^T,\I_N^T,\bG_{\rm CS}^T]^T$. The sub-matrices $\bG_{\rm W}$ and $\bG_{\rm CS}$ are $\frac{N_{\rm w}}{2} \times N$ matrices which consist of the last and the first $\frac{N_{\rm w}}{2}$ rows of the identity matrix $\I_N$, respectively. ${\bf{H}}^{(i)}_{\rm t}$ is the $N{\times}N$ circulant channel matrix of user $i$ with the first column being ${\bf h}^{(i)}$ zero padded to have the length of $N$. The windowed and aliased signal can be written as
\be\label{eqn:r_windowed}
\br^\prime=\sum^{K}_{i=1}{{e^{{\frac{j2{\pi}{\epsilon_i}N_{\rm GI}}{N}}}}(\bar{\bT}^T\bW{{\widetilde{\bPhi}}(\epsilon_i)\bar{\bT}){\bf{H}}^{(i)}_{\rm t}{\bf{s}}^{(i)}_{\rm t}}}+\bar{\bT}^T\bW\bR_{\rm GI}{\boldsymbol{\nu}},
\ee
where $\bW={\rm diag}({\bf w}_{\rm rc})$, ${\bf w}_{\rm rc}$ is the raised-cosine window vector and $\bar{\bT}^T$ does the aliasing operation as mentioned earlier. Since, ${\bf{H}}^{(i)}_{\rm t}$ is a circulant matrix, it can be spectrally factorized as ${\bf F}^{H}_{N}{\bf{H}}^{(i)}_{\rm f}{\bf F}_{N}$ where ${\bf{H}}^{(i)}_{\rm f}$ is the $N{\times}N$ diagonal matrix whose diagonal elements are the channel frequency response of the user $i$. Hence, recalling (\ref{eqn:si}), ${\bf{H}}^{(i)}_{\rm t}{\bf{s}}^{(i)}_{\rm t}$ can be written as ${{\bf F}^H_{N}}{\bf{H}}^{(i)}_{\rm f}{\bf{s}}^{(i)}_{\rm f}$ and after passing the signal through the DFT block, we have
\begin{eqnarray}\label{eqn:rdft}
\bar{\br}&=&{{\bf F}_{N}}{\br^\prime}\nonumber\\
&=&\sum^{K}_{i=1}{{e^{{\frac{j2{\pi}{\epsilon_i}N_{\rm GI}}{N}}}}({{\bf F}_{N}}\bar{\bT}^T\bW{{\widetilde{\bPhi}}(\epsilon_i)\bar{\bT}{{\bf F}^H_{N}}){\bf{H}}^{(i)}_{\rm f}{\bf{s}}^{(i)}_{\rm f}}}+{\tilde{\boldsymbol{\nu}}}\nonumber\\
&=&{\Lamb}{\bf x}+{\tilde{\boldsymbol{\nu}}},
\end{eqnarray}
where
\be\label{eqn:x}
{\x} = \sum^{K}_{i=1}{e^{{\frac{j2{\pi}{\epsilon_i}N_{\rm GI}}{N}}}}{\bf{H}}^{(i)}_{\rm f}{\bf{s}}^{(i)}_{\rm f}={\bf\bar{H}}_{\rm f}{\bf\bar{d}},
\ee
and
\be\label{eqn:lambda}
{\Lamb}=\sum^{K}_{i=1}{{\bf F}_{N}}\bar{\bT}^T\bW{\widetilde{\bPhi}}(\epsilon_i)\bar{\bT}{{\bf F}^H_{N}}{\boldsymbol\Pi}^{(i)},
\ee
is the $N{\times}N$ interference matrix. The matrix ${\boldsymbol\Pi}^{(i)}={\boldsymbol\Gamma}^{(i)}{({\boldsymbol\Gamma^{(i)}})^H}$ and ${\tilde{\boldsymbol{\nu}}} = {{\bf F}_{N}}\bar{\bT}^T\bW\bR_{\rm GI}{\boldsymbol{\nu}}$. The $N{\times}N$ diagonal matrix ${\bf\bar{H}}_{\rm f}$ contains the composite channel frequency response of all the users in its diagonal elements. It is worth mentioning that the phase factors, ${e^{{\frac{j2{\pi}{\epsilon_i}N_{\rm GI}}{N}}}}$, are absorbed into the composite channel of the users. The composite data vector ${\bf\bar{d}}$ includes the information symbols of all the users corresponding to their allocated subcarriers as if there has been no interference.

\begin{figure}[t]
		\centering
		\subfigure[]{
    \includegraphics[scale=0.43]{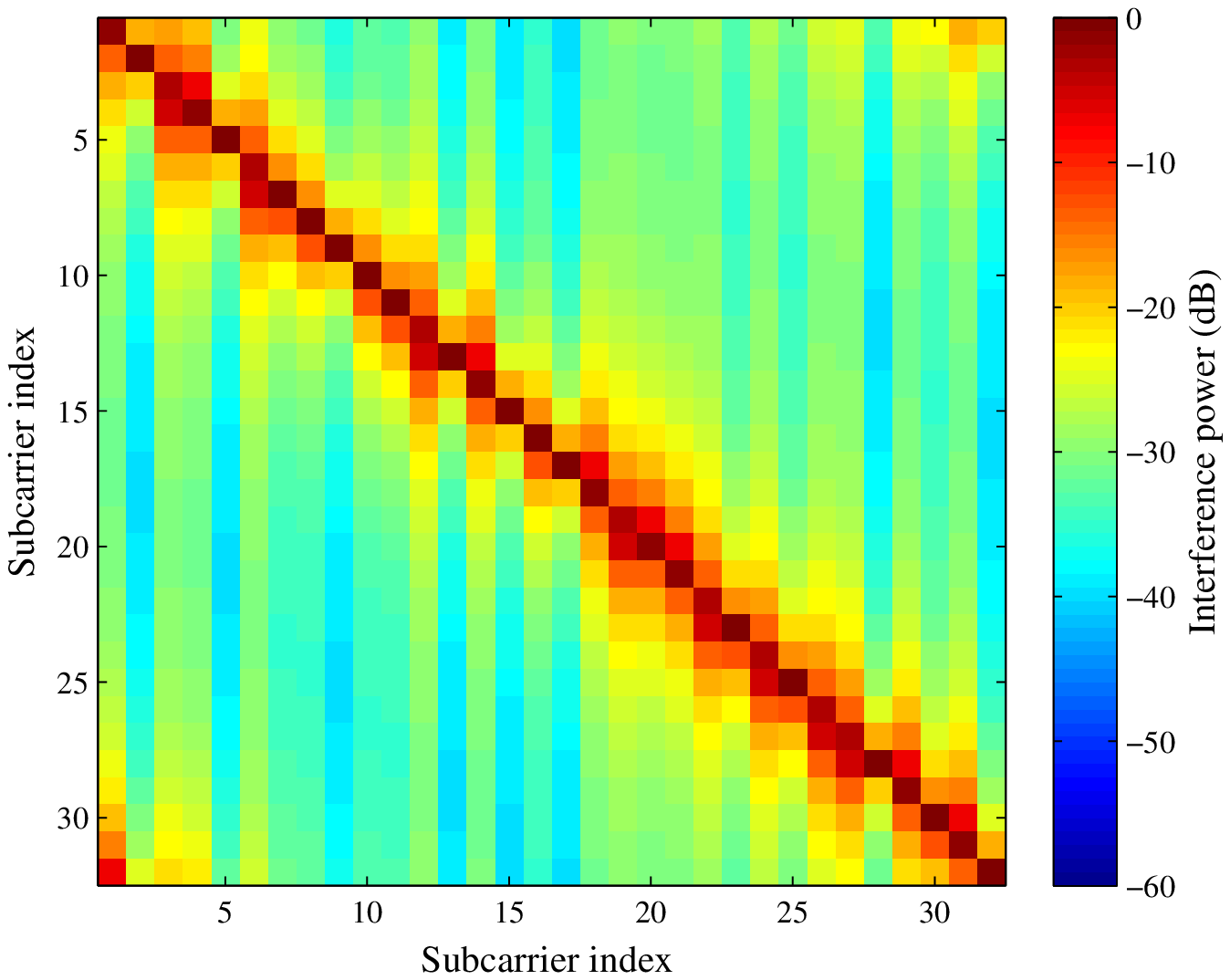}
    \label{fig:MAI2}
    }
	\subfigure[]{
    \includegraphics[scale=0.43]{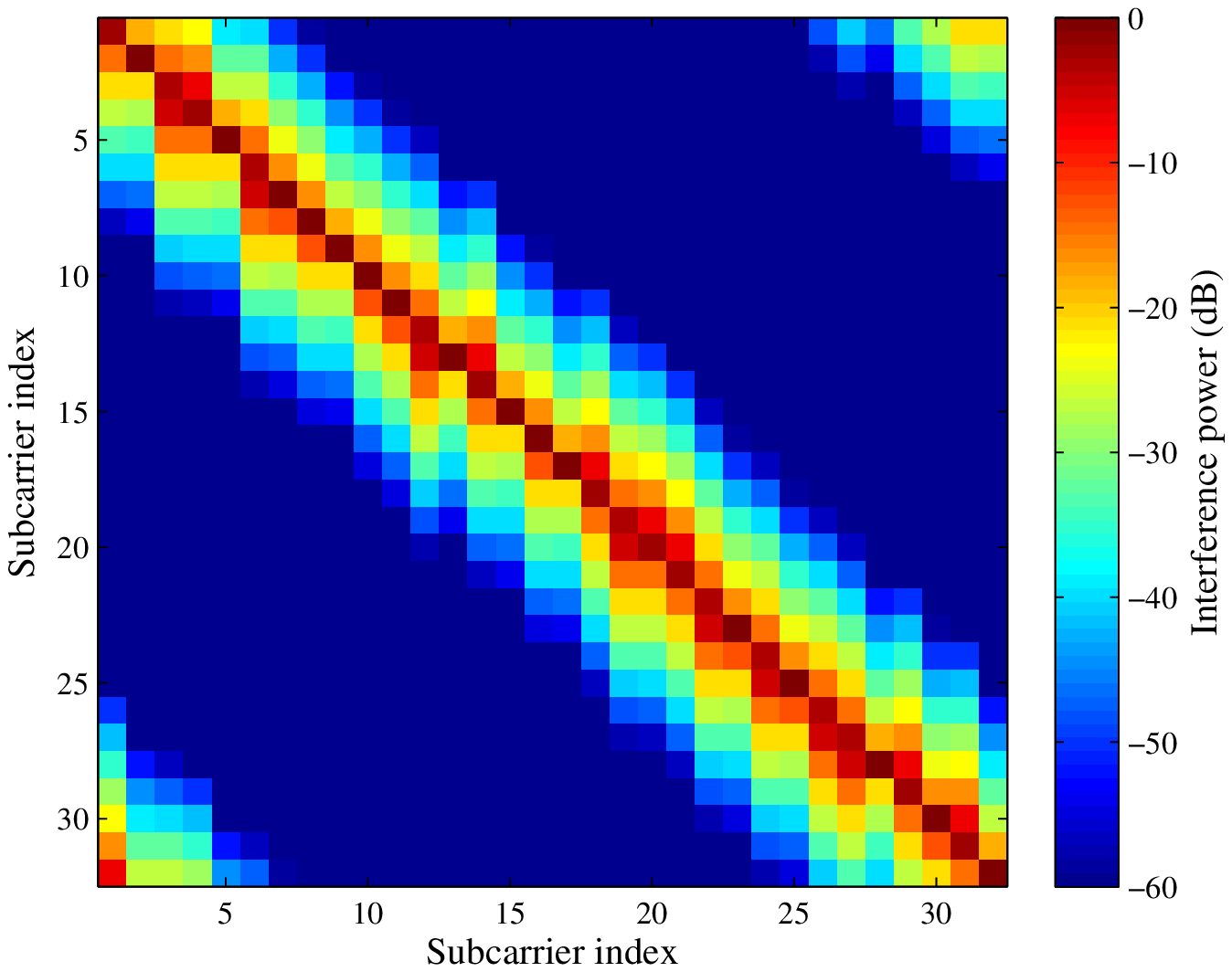}
    \label{fig:MAI1}
	}
\caption{Interference power between different subcarriers for G-CAS case when $N=32,~K=4,~N_{\rm w}=8,~{\rm CFOs}=[0.20,-0.35,0.45,-0.11]$ \subref{fig:MAI2} shows the interference power without receiver windowing \subref{fig:MAI1} shows interference power for OFDMA system with windowing at the receiver.}
\label{fig:MAI}
\vspace{-2 mm}
\end{figure}

As mentioned earlier, the effect of receiver windowing is that it limits the interference generated by each subcarrier on its adjacent ones within a certain distance which depends on the roll-off factor of the window. Fig. \ref{fig:MAI}, shows the quantized value of the interference powers among different subcarriers for both cases of OFDMA system with and without receiver windowing. In Fig.~\ref{fig:MAI}, we consider an OFDMA system with $32$ subcarriers and $4$ users with G-CAS. The CFOs are randomly chosen from a uniform distribution in the range $(-0.5,0.5]$. From Fig. \ref{fig:MAI}. \subref{fig:MAI1}, one may notice that the interference power in the system with receiver windowing after a certain band is very small and hence negligible. Thus, the interference matrix can be approximated with a quasi-banded matrix with very good precision and the simulation results which will be presented in Section \ref{sec:Sim_results} will attest to this fact. It can be inferred from Fig. \ref{fig:MAI}. \subref{fig:MAI2} that approximation of the interference matrix with a banded matrix as is proposed in \cite{Cao2007} while having large CFOs will result in a significant performance loss \cite{Lee2012}. 

Based on the above discussion, the interference matrix $\Lamb=\Lamb_{\rm QB}+\Lamb_{\rm I}$ where $\Lamb_{\rm QB}$ is the quasi-banded part and $\Lamb_{\rm I}$ contains the negligible part of the matrix. 
\be\label{eqn:Lambda_QB}
\left[\Lamb_{\rm QB}\right]_{m,n}=
\left\{
	\begin{array}{ll}
	[\Lamb]_{m,n}, &|m-n|<D, \\
	{}&|m-n|>N-D, \\
	0, & \rm{otherwise},
	\end{array}
\right. 
\ee
where the bandwidth of the matrix $\Lamb_{\rm QB}$ is $2D+1$.

Assuming the perfect knowledge about the CFOs at the base station and approximating $\Lamb$ with $\Lamb_{\rm QB}$, transmitted symbols of the users can be estimated using a zero forcing (ZF) technique as
\be\label{eqn:ZF}
{\hat{{\x}}}_{\rm{ZF}}=\Lamb_{\rm QB}^{-1}{\bar{\br}}=\x+{\Lamb_{\rm QB}^{-1}}{\Lamb_{\rm I}}\x+{\Lamb_{\rm QB}^{-1}}{\tilde{\boldsymbol{\nu}}}.
\ee

As long as the elements in ${\Lamb_{\rm I}}$ are very small, they do not have any impact on the precision of the detected data symbols. Using structural properties of the quasi-banded matrices, the ZF technique can be efficiently implemented. The interference matrix ${\Lamb_{\rm QB}}$ can be factorized into lower and upper triangular matrices, i.e., ${\Lamb_{\rm QB}}=\bL\bU$ using LU factorization. Since ${\Lamb_{\rm QB}}$ is a quasi-banded matrix, the matrices $\bL$ and $\bU$ inherit some kind of banded properties; thereby, they have many zeros and data symbol detection can be implemented with low computational complexity by using forward and backward substitution techniques. 

\section{Simulation Results}\label{sec:Sim_results}
In this section, we will compare the bit error rate (BER) performance of our technique for uplink OFDMA system with the solutions proposed in \cite{Cao2007} and \cite{Lee2012}. In the present solution we make an improvement to the ZF technique in \cite{Cao2007} by making use of windowing at the receiver. We also compare the BER of our technique with the one in \cite{Lee2012}, which coincides with the optimal MMSE solution.
 
In our simulations we consider an uplink system with $N=128$ subcarriers that are allocated to $K=4$ users based on G-CAS. Therefore, the total number of subcarriers assigned to each user is $L=32$. We assume that the users are using uncoded $4$-QAM modulation and their signals have gone through the SUI-2 channel proposed by the IEEE802.16 broadband wireless access working group \cite{SUI}. It is worth mentioning that the users experience independent multipath channels. The normalized CFOs are randomly chosen from a uniform distribution within the range $(-0.5,0.5]$ and we have $10000$ simulation runs. The receiver window that is used in our simulations is a raised-cosine window with $N_{\rm w}=14$ and the width $D=10$ is considered for both banded and quasi-banded approximation of the interference matrix. In the banded approximation of the interference matrix, we do not consider the top right and bottom left elements of the matrix in multiple CFO compensation process.

\begin{figure}[t!]
		\centering
    \includegraphics[scale=0.52]{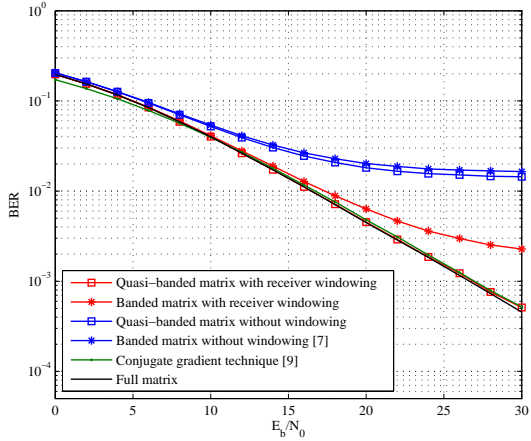}
    \caption{BER performance comparison of uplink OFDMA systems with and without receiver windowing when users are using G-CAS and normalized CFOs are $[-0.44,0.09,-0.34,0.18]$.}
\label{fig:BER}
\vspace{-3 mm}
\end{figure}

Banded approximation of the interference matrix without any filtering at the receiver side is suggested in \cite{Cao2007}. However, approximation of the interference matrix with a banded one is not very precise and, as shown in Fig.~\ref{fig:BER}, the BER performance is greatly degraded compared to ZF solution using the full interference matrix, as testified by the error floor of the starred blue curve. From Fig.~\ref{fig:MAI}, we can see that there is interference power in the top right and bottom left parts of the matrix. This may lead to the conclusion that using a quasi-banded rather than a banded interference matrix may result in better performance. However, as Fig.~\ref{fig:BER} shows, the performance improvement is small.

The addition of windowing at the receiver is much more effective. In the case of the banded matrix there is significant improvement in performance as Fig.~\ref{fig:BER} shows. The performance can be further improved when a quasi-banded matrix is used and in fact it approaches the case of the ZF solution using the full interference matrix.

For the purposes of valid comparison, it should be noted that the quasi-banded technique with receiver windowing also performs as well as the CG technique in \cite{Lee2012} but with the advantage of using a much simpler type of receiver.

\makeatletter
\setlength{\@fptop}{0pt}
\makeatother

\section{Experimental Study}\label{sec:Experimental_Study}
In this section, we present some results from an experimental wireless setup. Our system setup consists of a pair of National Instruments (NI) software defined radios. Each radio consists of a controller made up of an NI-PXIe-1082 and an NI-PXIe-1071 for transmitting and receiving the users' data, respectively, through an NI Flex-RIO FPGA module, NI-PXIe-7965R. This FPGA module is equipped with a Xilinx Vertex-5 FPGA. The FPGA sends and receives data to and from an NI RF Transceiver Adapter Module, NI 5791, over the air. On the transmitter side, the Transceiver Adapter Module, effectively, takes a digital signal from the FPGA and converts it to an analog signal through a $16$-bit digital-to-analog converter (DAC) that operates at $130$ MSamples/s, and modulates the result to a radio frequency (RF) carrier. The RF carrier frequency can take any value in the range of $200$ MHz through $4.4$ GHz and may have a bandwidth of up to $100$ MHz. On the receiver side, the Transceiver Adapter Module uses an analog-to-digital converter (ADC) that also operates at $130$ MSamples/s, and provides a resolution of $14$ bits/sample. 

We consider the case where four mobile users transmit simultaneously to a base station. However, since our hardware is limited to two communicating radios only, we have taken the following approach to emulate the presence of four users. The transmitting radio transmits the signal corresponding to each user, one at a time. The signals of the four users that experience different channels (as we have changed the position of transmitting radio for each user) are captured by the receiver and added together, after adding a controlled CFO for each. It is worth mentioning that all the received signals from our four users have the same signal to noise ratios (SNRs). The SNR of the uplink signal in our experiment is approximately $25~\rm{dB}$. Data symbols are from a $16$-QAM alphabet. To obtain an average performance, $500$ OFDM data packets are transmitted, each consisting of a preamble (for synchronization and channel estimation) followed by $6$ OFDM data symbols. The generalized carrier assignment scheme has been used to transmit the data of the four users over a bandwidth of $16.25$ MHz and across $512$ subcarriers. $126$ subcarriers are allocated to each user. The zero-th subcarrier is not used as our hardware adds an undesirable DC offset to it. Also, we have some guard-band subcarriers. Hence, the total number of active subcarriers is $504~(=126\times4)$. 

\begin{figure}[t]
		\centering
		\subfigure[]{
    \includegraphics[scale=0.42]{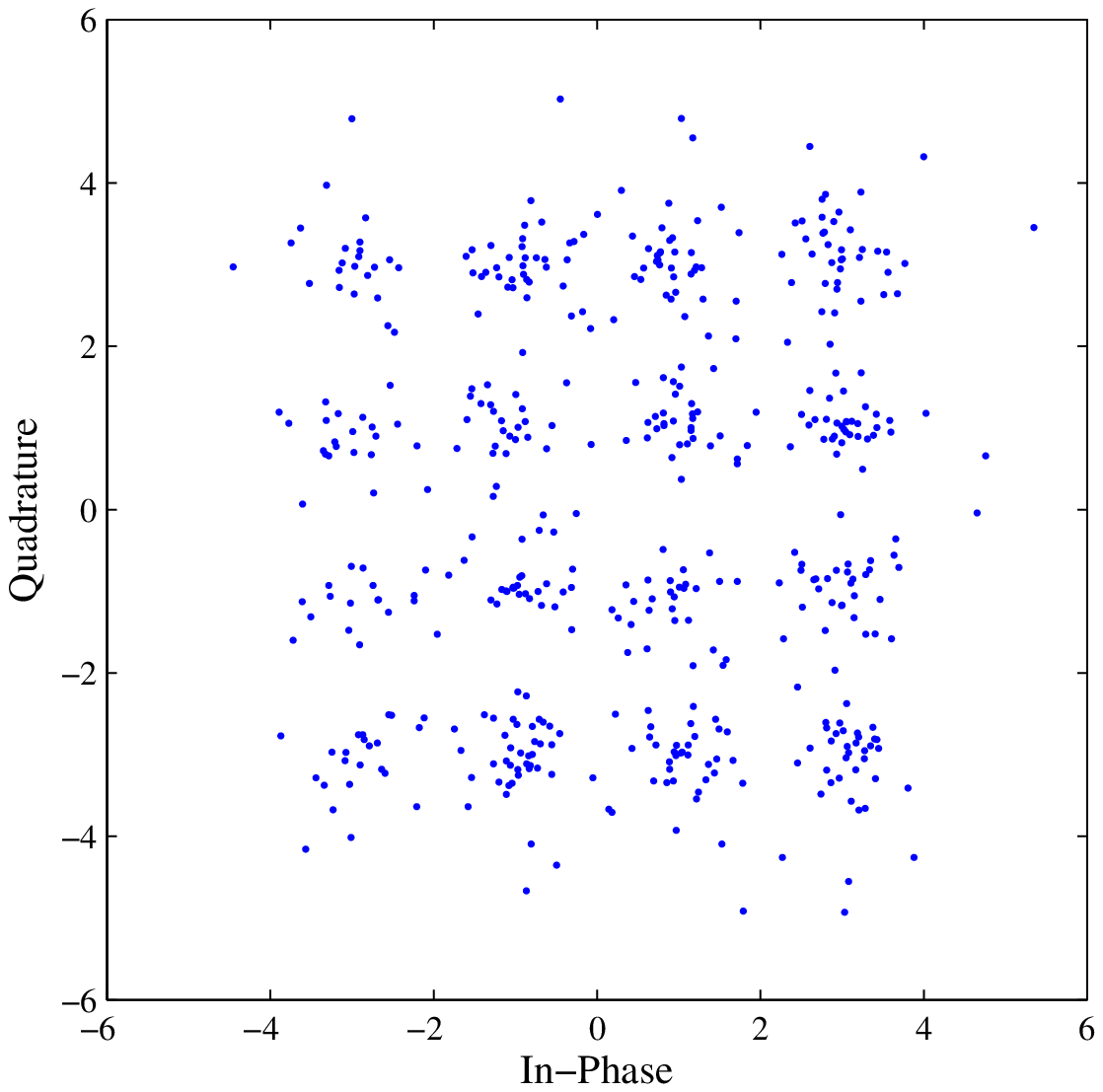}
    \label{fig:Tureli_Const}
    }
	\subfigure[]{
    \includegraphics[scale=0.42]{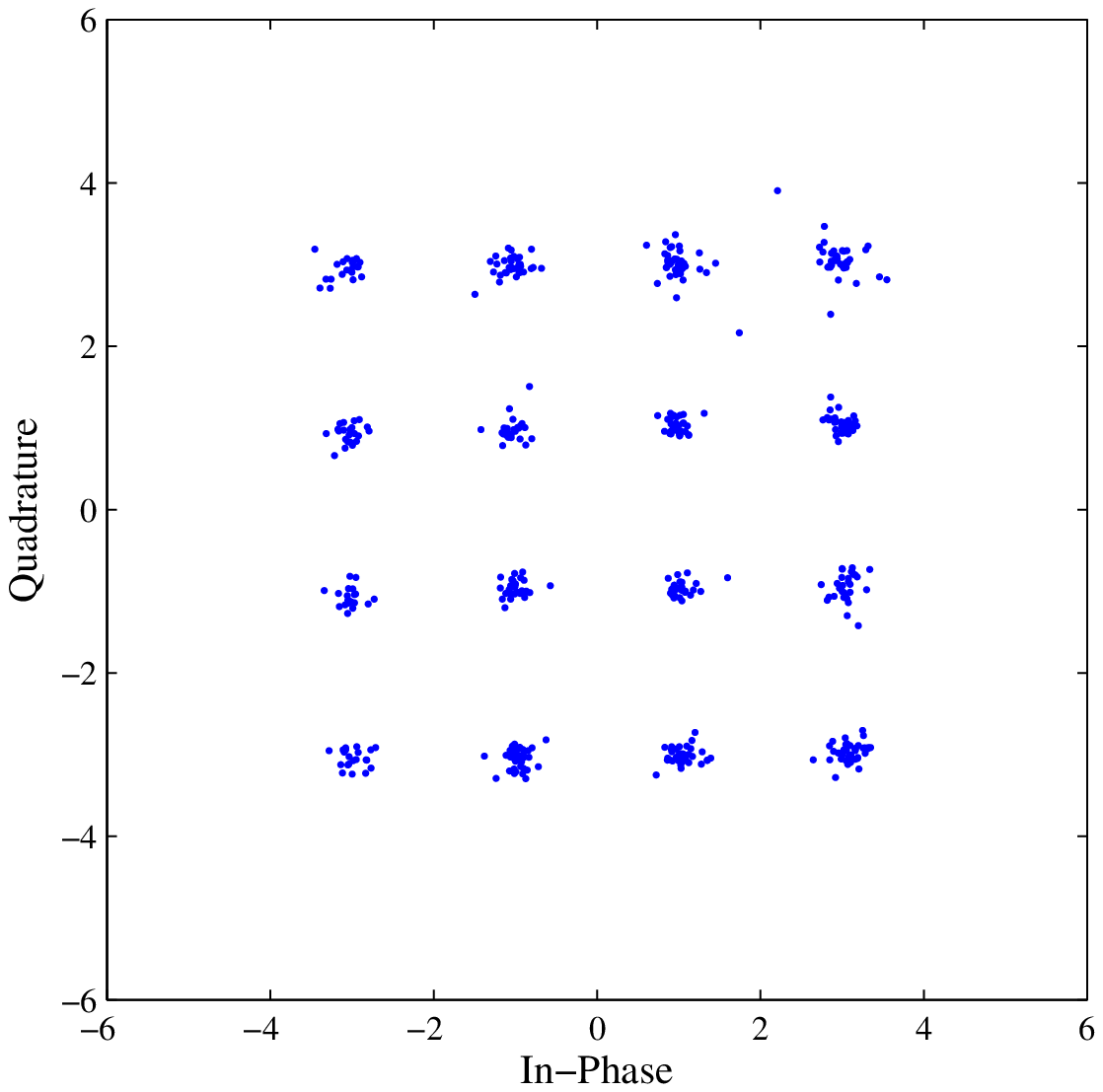}
    \label{fig:FOFDMA_Const}
	}
\caption{The scattered plots of \subref{fig:Tureli_Const} the proposed method in \cite{Cao2007} and \subref{fig:FOFDMA_Const} our proposed method.}
\label{fig:Constellations}
\vspace{-5 mm}
\end{figure}

Our method and that of \cite{Cao2007} are applied to the received signal (the combination of the four users' signals), to obtain the recovered data symbols according to (\ref{eqn:ZF}). These soft values are subtracted from the transmitted symbols to obtain the residual error due to the channel noise as well as any other effect due to the wireless channel. The ratio of the data symbol powers over the residual error provides an estimate of signal-to-interference plus noise ratio (SINR). The measured SINR for our method turns out to be $24.7~\rm{dB}$, while that of \cite{Cao2007} was only $16.4~\rm{dB}$. That is, the method proposed in this paper performs over $8~\rm{dB}$ better than the method of \cite{Cao2007} on average. These numbers and the fact that the SNR at the input is around $25~\rm{dB}$ indicate that while the method of this paper suffers from a negligible level of ICI, the less sophisticated method of \cite{Cao2007} is dominated by a significant level of ICI. This clearly is consistent with our simulation results in the previous section, where we found the method of \cite{Cao2007} suffers from a BER floor. To further show the superior performance of the method of this paper over that of \cite{Cao2007}, an example of the scattered plots of the two methods are presented in Fig. \ref{fig:Constellations}.

\section{System Complexity}\label{sec:sys_comp}
Table~\ref{tab:1} summarizes the computational complexity of our technique, the CG technique and the direct ZF technique which involves direct computation of the inverse of the full interference matrix $\Lamb$. It is assumed that $N$ subcarriers are allocated to $K$ users based on G-CAS. Parameter $I$ in the table indicates the number of iterations in the CG algorithm. Since all the operations include complex numbers, the complexity expressions are presented based on the number of complex multiplications (CMs). Compared to \cite{Cao2007}, the complexity of our method is slightly higher.

The conjugate gradient solution offers a good computational complexity reduction for the case of G-CAS with respect to the other existing solutions \cite{Lee2012}. However, the computational complexity of the CG technique depends on the number of users while our proposed technique is independent of the number of users. Therefore, in some cases we are able to reduce the complexity by over an order of magnitude. As a case in point, for $N=512$, $D=10$, $K=8$ and $I=32$ (as suggested in \cite{Lee2012}), complexity of the proposed technique with quasi-banded matrix approximation is only $16.5\%$ of the CG technique. As the number of users increases from $8$ to $16$, the computational complexity of our proposed technique is only $9\%$ of the CG method. Hence, our technique is more computationally efficient than the CG algorithm as the number of users increases. Another point to make here is that since in (\ref{eqn:Lambda_QB}), we have many zero elements, the matrix $\Lamb_{\rm QB}$ is a sparse matrix. Therefore, sparse storage techniques can be applied to significantly reduce the memory requirements.

\renewcommand{\arraystretch}{1.5}
\begin{table}[t]
  \centering
    \caption{Computational Cost of Different CFO Compensation Techniques}
    \label{tab:1}
    \resizebox{0.48\textwidth}{!}
{\small \begin{tabular}{|c|c|}
\hline\hline
Technique & Number of Complex Multiplications \\ \hline\hline
Direct ZF & $\frac{1}{3}N^3+2N^2+\frac{KN}{2}\log_2N$ \\ \hline
CG, \cite{Lee2011} & $I(KN\log_2N+2KN+5N)+KN\log_2N+2KN$ \\ \hline

Quasi-banded matrix & $(4N-10)D^2+8ND-\frac{16}{3}D^3-\frac{11}{3}D+\frac{KN}{2}\log_2N+N_{\rm w}$ \\ \hline\hline
    \end{tabular}}
\vspace{-2 mm}
\end{table}

\section{Conclusion}\label{sec:Conclusion}
In this paper, we proposed a novel method to reduce the CFO induced interference in the uplink of OFDMA systems. Our method is applicable to the generalized carrier allocation scheme. We suggested the use of receiver windowing for interference suppression; an idea that we borrowed from DSL literature. The interference power in OFDMA system with windowing is more concentrated towards the main diagonal of the interference matrix compared to the conventional OFDMA system. This interference localization comes at the expense of a longer cyclic extension, but with the gain of a simpler receiver. We have shown that after the application of windowing at the receiver, the interference matrix can be approximated well with a quasi-banded matrix while keeping very close to the optimal performance. A substantial amount of saving in computational complexity has been achieved using our proposed solution which makes it attractive for hardware implementations. Simulation and experimental results that compare the advantageous trade-offs of the proposed method with the existing methods were also presented.

\bibliographystyle{IEEEtran}

\begin{thebibliography}{10}
\providecommand{\url}[1]{#1}
\csname url@samestyle\endcsname
\providecommand{\newblock}{\relax}
\providecommand{\bibinfo}[2]{#2}
\providecommand{\BIBentrySTDinterwordspacing}{\spaceskip=0pt\relax}
\providecommand{\BIBentryALTinterwordstretchfactor}{4}
\providecommand{\BIBentryALTinterwordspacing}{\spaceskip=\fontdimen2\font plus
\BIBentryALTinterwordstretchfactor\fontdimen3\font minus
  \fontdimen4\font\relax}
\providecommand{\BIBforeignlanguage}[2]{{%
\expandafter\ifx\csname l@#1\endcsname\relax
\typeout{** WARNING: IEEEtran.bst: No hyphenation pattern has been}%
\typeout{** loaded for the language `#1'. Using the pattern for}%
\typeout{** the default language instead.}%
\else
\language=\csname l@#1\endcsname
\fi
#2}}
\providecommand{\BIBdecl}{\relax}
\BIBdecl

\bibitem{Chockalingam2009}
K.~Raghunath and A.~Chockalingam, ``{SIR analysis and interference cancellation
  in uplink OFDMA with large carrier frequency/timing offsets},'' \emph{IEEE
  Trans. Wireless Commun.}, vol. 8, no.5, pp. 2202--2208, May 2009.

\bibitem{Moreli2007}
M.~Morelli, C.~C.~J. Kuo, and M.~O. Pun, ``{Synchronization techniques for
  orthogonal frequency division multiple access (OFDMA): A tutorial review},''
  \emph{Proc. of the IEEE}, vol. 95, no.7, pp. 1394--1427, July 2007.

\bibitem{Linesearch2007}
Y.~Na and H.~Minn, ``{Line search based iterative joint estimation of channels
  and frequency offsets for uplink OFDM systems},'' \emph{IEEE Trans. Wireless
  Commun.}, vol.~6, no.~12, pp. 4374--4382, 2007.

\bibitem{Ghrayeb2009}
X.~N. Zeng and A.~Ghrayeb, ``Joint cfo and channel estimation for ofdma uplink:
  an application of the variable projection method,'' \emph{IEEE Trans.
  Wireless Commun.}, vol.~8, no.~5, pp. 2306--2311, 2009.

\bibitem{Arsalan2007}
T.~Yucek and H.~Arslan, ``{Carrier frequency offset compensation with
  successive cancellation in Uplink OFDMA Systems},'' \emph{IEEE Trans.
  Wireless Commun.}, vol. 6, no.10, pp. 3546--3551, Oct. 2007.

\bibitem{Lataief2005}
H.~Defeng and K.~B. Letaief, ``{An interference-cancellation scheme for carrier
  frequency offsets correction in OFDMA systems},'' \emph{IEEE Trans. Commun.},
  vol. 53, no.7, pp. 1155--1165, July 2005.

\bibitem{Cao2007}
Z.~Cao, U.~Tureli, and Y.-D. Yao, ``{Low-complexity orthogonal spectral signal
  construction for generalized OFDMA uplink with frequency synchronization
  errors},'' \emph{IEEE Trans. Veh. Technol.}, vol. 56, no. 3, pp. 1143--1154,
  May 2007.

\bibitem{Hsu2008}
C.~Y. Hsu and W.~R. Wu, ``{A low-complexity zero-forcing CFO compensation
  scheme for OFDMA uplink systems},'' \emph{IEEE Trans. Wireless Commun.}, vol.
  7, no.10, pp. 3657--3661, Oct. 2008.

\bibitem{Lee2011}
K.~Lee and I.~Lee, ``{CFO compensation for uplink OFDMA systems with conjugated
  gradient},'' \emph{Proc. of the IEEE ICC '11}, pp. 1--5, June 2011.

\bibitem{Lee2012}
K.~Lee, S.-R. Lee, S.-H. Moon, and I.~Lee, ``{MMSE-based CFO compensation for
  uplink OFDMA systems with conjugate gradient},'' \emph{IEEE Trans. Wireless
  Commun.}, vol. 11, no.8, pp. 2767--2775, Aug. 2012.

\bibitem{S.Ahmed2009}
S.~Ahmed and L.~Zhang, ``{Low complexity iterative detection for OFDMA uplink
  with frequency offsets},'' \emph{IEEE Trans. Wireless Commun.}, vol.~8,
  no.~3, pp. 1199--1205, 2009.

\bibitem{AF2013}
A.~Farhang, N.~Marchetti, and L.~Doyle, ``{Low Complexity LS and MMSE Based CFO
  Compensation Techniques for the Uplink of OFDMA Systems},'' \emph{Proc. of
  the IEEE ICC '13}, June 2013.

\bibitem{Chen2010}
G.~Chen, Y.~Zhu, and K.~B. Letaief, ``{Combined MMSE-FDE and interference
  cancellation for uplink SC-FDMA with carrier frequency offsets},''
  \emph{Proc. of the IEEE ICC '10}, pp. 1--5, May 2010.

\bibitem{Schniter2004}
P.~Schniter, ``{Low-complexity equalization of OFDM in doubly selective
  channels},'' \emph{IEEE Trans. Signal Processing.}, vol.~52, no.~4, pp.
  1002--1011, 2004.

\bibitem{Zipper}
F.~Sjoberg, R.~Nilsson, M.~Isaksson, P.~Odling, and P.~Borjesson,
  ``{Asynchronous Zipper},'' \emph{Proc. of the IEEE ICC '99}, pp. 231--235
  vol.1, 1999.

\bibitem{FarhangOFDMvsFBMC}
B.~Farhang-Boroujeny, ``{OFDM Versus Filter Bank Multicarrier},'' \emph{Signal
  Processing Magazine, IEEE}, vol.~28, no.~3, pp. 92--112, 2011.

\bibitem{SUI}
\emph{\rm{The IEEE802.16 Broadband Wireless Access Working Group, Channel
  Models for Fixed Wireless Applications [Online]. Available:
  http://www.ieee802.org/16/tg3/contrib/802163c-01\_29r4.pdf}}.

\end{thebibliography}

\end{document}